# Unleashing the Power of Hashtags in Tweet Analytics with Distributed Framework on Apache Storm

Vibhuti Gupta and Rattikorn Hewett
Department of Computer Science
Texas Tech University, Lubbock, TX 79415
Emails: vibhuti.gupta@ttu.edu, rattikorn.hewett@ttu.edu

*Abstract*—**Twitter is a popular social network platform where users can interact and post texts of up to 280 characters called** *tweets*. **Hashtags, hyperlinked words in tweets, have increasingly become crucial for tweet retrieval and search. Using hashtags for tweet topic classification is a challenging problem because of context dependent among words, slangs, abbreviation and emoticons in a short tweet along with evolving use of hashtags. Since Twitter generates millions of tweets daily, tweet analytics is a fundamental problem of Big data stream that often requires a real-time Distributed processing. This paper proposes a distributed online approach to tweet topic classification with hashtags. Being implemented on** *Apache Storm*, **a distributed real time framework, our approach incrementally identifies and updates a set of strong predictors in the Naïve Bayes model for classifying each incoming tweet instance. Preliminary experiments show promising results with up to 97% accuracy and 37% increase in throughput on eight processors.**

## I. Introduction

The proliferation of social media networks in last few years have produced an enormous volumes of data and become a common source of Big data. Twitter is one of the most popular social media platform, where users post short text messages of up to 280 characters, known as *tweets* for communication. On average, 6000 tweets are generated per second and 500 million tweets per day. Since twitter generates huge, unstoppable, fast growing and unstructured Big data stream of tweets daily, tweet analytics is a fundamental problem of Big data stream that often requires real-time Distributed processing.

Hashtags, user-defined hyperlinked words of typical topics, in tweets facilitate efficient information sharing [14]. Hashtags begin with a hash symbol representing various subjects, for examples, #election, #happy, #partying, #nba, #Oscars2016 conveys a topic, emotion, action, official organization, or event, respectively. They are crucial for trend/event detection, search/retrieval and advertisement. Hashtags have been adopted and quickly become common in many blogging sites and social media platforms including Facebook, Instagram, Flickr, Tumblr and Pinterest.

Recent research in tweet analytics has studied how hashtags can be effectively applied [2, 3, 6, 11-14, 17]. While many of hashtag applications are successful, tweet classification remains challenging largely due to the nature of tweets and hashtags whose trends can quickly evolve. Tweets have a limited number of words making it hard to derive contexts from dependent words. We also have to cope with ambiguity, slangs, abbreviations and emoticons in tweets. To make things worse, there is no standard on how hashtags are created or expressed. The same subject can have different hashtags defined by different users (e.g., #omg, #ohmygod). Majority of tweet classification [3, 12, 13, 14] deals with sentiment analysis where sentiment classes can be described by semantic of keywords or hashtags while a topic requires a diverse set of hashtags to cover various aspects of it.

Our recent work [7] introduced a *hybrid hashtag* approach to cope with the challenges of tweet topic classification using hashtags. *Hybrid Hashtags* consist of two types of hashtags: 1) those that are extracted from input tweet data and 2) those derived from a knowledge base of topic (or class) concepts (or topic ontology) by using *hashtagify* [18], a tool to generate "similar" hashtags from a given term (see more details in [7]). We evaluated the effectiveness of this semi-automated approach using a batch analysis on Naïve Bayes algorithm. The applicability of this approach in real tweet Big data stream requires an online and distributed approach to deal with fast and dynamic arrival rates of tweets. Thus, real time processing with minimum latency is desirable.

This paper is different from our previous work [7] in that it presents a fully automated, online and distributed system for tweet topic classification using *Hybrid Hashtags* as opposed to finding the most effective way to use hashtags for tweet classification in a non-distributed environment. Our contribution is two fold in this paper. First, we propose an online approach (both for data pre-processing and analytic) to analyzing each tweet to identify appropriate *hybrid hashtags* and incrementally updating an accumulated set of hybrid hashtags. Our approach is completely online since it updates the hybrid hashtags and classification model incrementally with each incoming tweet instance. Second, we empirically illustrate how the proposed approach is scalable and thus, suitable for Big data stream environment giving a lower execution time and a higher throughput in Apache Storm framework. The rest of the paper is organized as follows. Section II discusses related work. Section III describes our proposed approach followed by experimental results in Section IV and Section V concludes the paper.

## II. RELATED WORK

Recent research in tweet analytics has studied how hashtags can be applied to tweet classification [2, 3, 12, 13, 14], tweet retrieval/search [5], and hashtag recommendation [6, 17].

Work on hashtag recommendation for a given tweet finds tweets similar to the given tweet and ranks hashtags in those tweets for recommendation based on how they are closely relevant to content of the given tweet using similarity measures, statistical model or probabilistic machine learning model [6, 17]. Most related hashtag research to our work is on tweet classification [3, 12, 13, 14], majority of which deals with sentiment analysis. There are some large scale implementations for twitter sentiment analysis [8,9]. [8] uses all hashtags and emoticons as sentiment labels and classify tweets using MapReduce and Apache Spark while [9] builds a large scale sentiment lexicon and classify using MapReduce. Previous work in [10] uses ontology to determine sentiment of twitter posts by assigning sentiment scores to each tweet instance and [16] uses ontology of keywords to classify the documents in economic field.

The difference between the above work and ours is not just by the domain but the characteristics of the classes to classify. Our previous work in [7] deals with tweet topic classification using domain specific knowledge and this paper extends it for Big data stream using Apache Storm [23]. The main distinction of our distributed processing with the previous approaches is that, our processing is completely online implemented on Storm framework as opposed to *MapReduce* [4], which is a distributed batch processing. Although both Apache Storm and Apache Spark are data stream processing, *Apache Storm* is an online framework while *Apache Spark* [22] processes data in batches and therefore not applicable to our work. However, combination of both can be applied for better computation on Stream data.

## III. PROPOSED APPROACH

The key distinction of this work is an online and distributed approach to a tweet analytic using *hybrid hashtags* as opposed to focusing on constructing a hybrid hashtag technique for a non-distributed computing on a single processor as proposed in our previous approach in [7]. In particular, we use *Naïve Bayes* [19] algorithm for classification. We now describe how we select and construct

| **Algorithm** *Hybrid-hashtags* ($H_C$, $H_T$, $k$) |
|---|
| **Inputs:** $H_C$, a set of ontology-driven hashtags corresponding class concepts; $H_T$, a set of hashtags extracted from tweet data sets; $k$, a set size of hashtags correlated with a given concept |
| **Output:** $H$, a set of hybrid hashtags (from $H_C$ and $H_T$) of $k$-correlation with $C$ |
| 01: $H_1 \leftarrow H_C \cap H_T$ // potentially strong predictive hashtags |
| 02: $H_2 \leftarrow \emptyset$ |
| 03: **For** each $h \in H_1$ **do** |
| 04:     $H \leftarrow Select(H_T - H_C, h, k)$ <br> // add new tweet hashtags that are $k$-correlated with $h$ |
| 05:     $H_2 \leftarrow H \cup H_2$ |
| 06: **end for** |
| 07: **return** $H_1 \cup H_2$ |

**Fig. 1** Combining tweet-extracted with ontology-driven hashtags.

appropriate hybrid hashtags [7] followed by our proposed online approach.

Our *hybrid hashtags* construction approach starts by building a domain-specific knowledge base describing concepts relevant to the class/topic to be classified. The knowledge base is a graphical representation of concepts and relationship between them (or synonymously referred to as *ontology*). Each of the concepts from the most bottom level nodes is fed into an automated tool *Hashtagify* [18] to retrieve a set of hashtags relevant to the input concept called *concept-based hashtags*.

Each of these concept-based hashtags is ranked according to its *correlation* with the given concept and a specified $k$. The correlation score between the hashtag $h$ and concept $c$ can be computed by using equation 1.

$$corr(c,h) = \frac{\sum_{i=1}^{n}(c_i - \bar{c})(h_i - \bar{h})}{(n-1)\ S_c S_h} \qquad (1)$$

where $c$ and $h$ are vectors representing frequency of occurrence of concept $c$ and hashtag $h$ in the *hashtagify* data while $\bar{c}$, $\bar{h}$, $S_c$, $S_h$ are the mean and standard deviation of values respectively. We extract top $k$ hashtags correlated with the concept (called *k-correlated* hashtags). This process repeats until we retrieve all the *k-correlated* hashtags of all selected concepts of a given class concept (or topic). These are *Ontology-driven hashtags*. This set of hashtags are combined with the *k-correlated tweet-based-hashtags* (i.e., top $k$ hashtags in the tweet data that are correlated with the class topic) to get the set of *hybrid hashtags*. Hybrid hashtag approach is shown in Figure 1. Detailed explanation can be found in [7]. These hybrid hashtags are used for tweet topic classification using various classification algorithms (i.e., Naïve Bayes, SVM, k-NN) to evaluate its performance. More details for the approach can be found at [7].

This paper presents an online and distributed approach to tweet topic classification using *Hybrid Hashtags*. The *hybrid hashtag* construction is online in the sense that the set of hybrid hashtags is updated incrementally for each new incoming tweet instance. The approach starts with an initial set of ontology-driven hashtags as developed in [7]. For each incoming tweet instance, each of hashtags in the tweet, compute its correlation with the topic concepts. Keep only the top $k$ tweet-based hashtags. Thus, we obtain an accumulated set of *k-correlated tweet-based-hashtags*. The latter set will grow as more tweets are analyzed. Combining this tweet-based hashtag set with the *ontology-driven-hashtags*, a set of *hybrid hashtags* for each tweet instance is obtained. These hybrid hashtags are used to classify the tweet using online Naïve Bayes classifier. Same process repeats for next tweet instance computing a new set of *hybrid hashtags* from current as well as previous tweet instances. In this way *hybrid hashtag* set updates itself with each new incoming tweet instance until the tweet stream ends. The classification model is updated with this new set of hybrid hashtags. In this way the classifier is incrementally updated and improved the classification results at each new instance of tweet until the stream ends.

The proposed approach has been implemented in *Apache Storm* [23], a distributed real-time processing framework for Big Data Streams. The Storm framework processes data in real time using spouts and bolts as components to make a topology. *Spouts* are the source of stream data that are being processed by *Bolts* to produce results. The topology is submitted to a *Master Node* known as *Nimbus* which distributes the computation among *worker nodes* in a cluster which executes a subset of specific topology running in its own JVM. Each worker node has multiple worker processes which executes the topology using *Executers*. Each worker process runs several *executors* and run in the worker's JVM. Each executor contains multiple *Tasks* that perform the actual data processing. The coordination between Master node and worker nodes is maintained by *Zookeeper*. Storm has been successfully applied for many data stream applications. A detailed description of Storm can be found in [23].

## IV. EXPERIMENTS AND RESULTS

This section provides the experiments and results to evaluate the effectiveness of the proposed approach in Big Data Stream Infrastructure. In this paper, the Storm cluster is composed by a varying number of virtual machines (VMs or processors) (i.e., 1, 2, 4, and 8) in a system with Intel Core -i7-8550U CPU 2 GHz processor, 16 GB RAM 8 cores and 1TB of Hard disk. Each of the virtual machine is configured with 4 vCPU and 4 GB RAM. We have installed Ubuntu 14.04.05 64 bits OS in each of the VM along with the JDK/JRE v 1.8. All nodes are running *Apache Storm* except the one running *Zookeeper* and *Nimbus* [23]. The *Apache Storm* version used is 0.9.7 with *zookeeper* 3.4.9.

To deploy our approach in Storm, we focus on using Naïve Bayes classifier on simulated dataset of tweets of size 8000 as in [7] with topic classes: entertainment and sport. For our experiments, five randomly selected keywords of each class are used as query words to Twitter API [20] for data collection. Each of the retrieved tweets is manually labeled in appropriate classes. Additional tweets of other classes are crawled, labeled and collected or discarded similarly. Many tweets are repeatedly retrieved so the process has to repeat many times before a certain data set size can be obtained.

The data stream is simulated by randomly repeating the stream of 8000 tweets. Therefore, the accuracy for our online classification is obtained from those of the first 8000 tweet instances.

**Table I**: Comparisons of Tweet Stream Classification Results.

| Approaches | Batch Analysis | Online Analysis |
|---|---|---|
| Words Only | 74.0 % | 77.0% |
| Words & Tweet Hashtags | 93.0 % | 95.0% |
| Tweet Hashtags Only | 89.0 % | 90.0% |
| **Our Hybrid Hashtags** | **95.0 %** | **97.0%** |

Table I compares the average accuracy results obtained by the previous batch analysis approach [7] with those obtained by the proposed online analysis approach using Naïve Bayes classification algorithm. The batch analysis attempts to find the best way to exploit hashtags in tweets for tweet classification. Thus, we explored the analysis in batches by investigating the power of hashtags compared to using only words in tweets.

As shown in Table I, like the batch analysis, hybrid hashtags give the best performing result (97% accuracy) compared to other online approaches using other set of features (i.e., words, words & tweet hashtags, etc.) Furthermore, in each set of features, the online preprocessing and classifier perform a little better than the batch analysis with 2% increase of accuracy in the hybrid hashtags. This could be due to the fact that the set of features for the online approach "gradually" adapts to learning from the growing input tweets as opposed to the fixed set of features pre-determined during the training of the batch analysis.

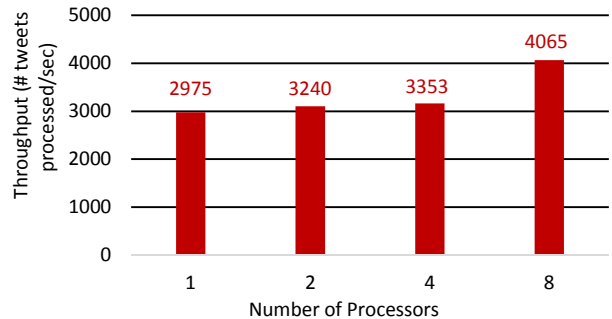

**Fig. 2** Throughput for hybrid hashtags with increasing number of processors

We also illustrate the throughput and processing time with *hybrid hashtags* approach to evaluate the scalability of the approach when the number of processors is increased. Throughput is a total number of tweet instances processed per unit time (i.e., seconds in our case) and processing time is the average time taken to completely process a tweet instance in the Storm architecture.

We ran the experiment for a session of 5 minutes in each of the case (i.e., with no. of processors as 1,2,4,8) and observed the throughput and execution time values. Figure 2 compares the throughput when the number of processors increases in the distributed environment. As shown in Figure 2, throughput improves from 2,975 tweets/sec to 4,065 tweets/sec when we increase from a single processor to eight processors.

Table II: % Increase in throughput with increasing number of processors

| Number of Processors | % Increase in throughput |
|---|---|
| 2 | 8.9% |
| 4 | 12.7% |
| 8 | **37.0%** |

Table II shows the percentage of increase in throughput with increasing number of processors as compared to a single processor. It shows a slight increase as we doubled and quadrupled the number of processors but reached highest of 37% increase when the number of processers is eight. The number of processed tweets depends upon the speed of execution, so the faster the tweets are processed, the higher the throughput is obtained. By increasing the number of processors, the rate of processing the tweets increases resulting in the improvement of the throughput for the classification.

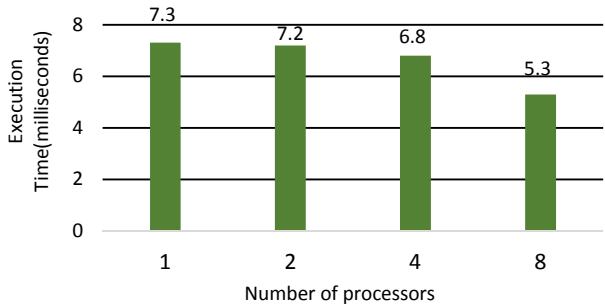

**Fig. 3** Processing time for hybrid hashtags with various processors

Figure 3 compares the execution times as the number of processors increases. It shows that the processing time is decreased on an increased number of processors since computational analysis of each tweet can now be distributed to multiple processors. Thus, the average execution time of each processor is decreased.

Table III shows the % reduction of execution time with increased number of processors as compared to single processor. As shown in Table III, there is a slight reduction in execution time when the number of processors is 2 and 4 but increases drastically with 8 processors.

Table III: % Reduction in execution time with various numbers of processors

| Number of Processors | % Reduction in Execution time (ms) |
|---|---|
| 2 | 1.4% |
| 4 | 6.8% |
| 8 | **27.0%** |

It is interesting to see that the % reduction of the time is not necessarily linear to the number of processors used. Further experiments are required.

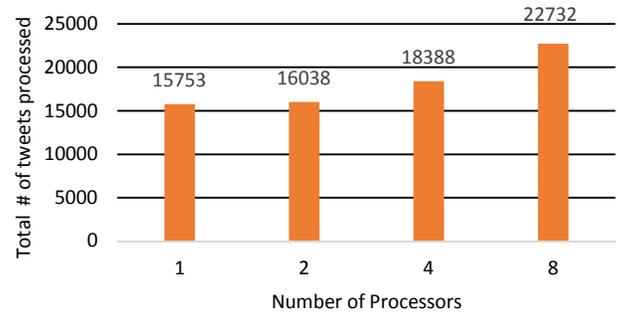

**Fig. 4** Total # of tweets being processed for various number of processors

Figure 4 shows the total number of tweets being processed at various number of processors. As the number of processors increases, the number of tweets is processed as expected. Thus, the proposed online approach appears to perform as well as expected. This shows promising applicability to Big data tweet analytics not only accuracy of the classification but also efficiency in data processing and analysis

## V. CONCLUSION

This paper presents an online, automated and distributed approach to use hybrid hashtags for tweet classification in Apache Storm. The approach is general in that it can be applied for any class concept in any domain. The experimental results show that the proposed approach is able to benefit from the distributed processing capabilities in reducing the execution time, scalability and providing real time data processing. Future work includes more experiments on different domains and applications of this online approach to other kinds of tweet analytics. Additional research using different windowing techniques are required to help improve tweet classification and other tweet analytic problems.